# Instant e-Teaching Framework Model for Live Online Teaching

Suhailan Safei[1], Mat Atar Mat Amin[2], Ahmad Nazari Mohd Rose[3] and Mohd Nordin Abdul Rahman[4]

Faculty of Informatics,
Universiti Sultan Zainal Abidin,
Gong Badak Campus,
21300 Kuala Terengganu,
Terengganu, Malaysia

**Abstract**
Instant e-Teaching is a new concept that supplements e-Teaching and e-Learning environment in providing a full and comprehensive modern education styles. The e-Learning technology depicts the concept of enabling self-learning among students on a certain subject using online reference and materials. While the instant e-teaching requires 'face-to-face' characteristic between teacher and student to simultaneously execute actions and gain instant responses. The word instant enhances the e-Teaching with the concept of real time teaching. The challenge to exercise online and instant teaching is not just merely relying on the technologies and system efficiency, but it needs to satisfy the usability and friendliness of the system as to replicate the traditional class environment during the deliveries of the class. For this purpose, an instant e-Teaching framework is been developed that will emulate a dedicated virtual classroom, and primarily designed for synchronous and live sharing of current teaching notes. The model has been demonstrated using a teaching Arabic recitation prototype and evaluated from the professional user profession's perspectives.

## 1. Introduction

Special courses for the public to either upgrade their skills or acquire new knowledge are currently been offered at most academic institutions. However, the diverse background of students interested in learning the offered courses by academic institutions has become a major hindrance for students to enroll themselves into those courses. It has become more complicated for the students who already have a career in their life, to attend those offered courses as a full-time basis. The main issue of this group of students is the period of their available time to attend those offered courses. Therefore to schedule for a fix time table that could cater the various free times of different students is almost impossible. It would therefore be more logical for them to be enrolled in this system of e-teaching whereby learning and teaching can be conducted at their own free time. By doing so, it would free them for disrupting their normal chores, since the e-teaching can be conducted outside their working time. Besides that, other factors such as the students' mobility will also make it difficult for the students to be continuously present in the classroom session. The mobility factor may be due to their work or health factors which will hinder them from attending the normal class consistently. As such, by enrolling into a distance learning program, their mobility elusiveness would not curtail them from learning the required skills.

[1] has suggested that there are three uses of information and communication technology (ICT) in a distance learning course. The first one is to make use of ICT is to support a resource-based learning approach where the students are given a wide choice of learning materials. Secondly, the use of ICT is to allow students to participate in virtual communication. Lastly, the third use of ICT is to promote an active approach to learning. It is therefore our intention here in this paper to highlight the integration of ICT technologies into our proposed model of Instant e-Teaching.

Distance learning courses or commonly known as e-learning, uses web technology as their main tool for their application. Such technology provides convenient ways to share resource-based learning materials amongst the users. Web-based communication systems have been widely advocated as tools for collaboration that can support self-explanation, social negotiation, and shared knowledge construction among participants [2].

Virtual communication enables users to instantly be in communication with each other via text, image, audio and video. Such features are commonly supported as Instant Messaging (IM) tools. IM actually refers to real time application software that enables electronic live communication among users using text, image, video and audio. Among the popular IMs are Yahoo Messenger, Skype, Windows Live Messenger, and AOL Messenger. This application concept can be used in distance learning to enable students or teachers to participate in virtual communication. It enhances real time teaching and discussion through live video/audio conference, instant command instructions and synchronous knowledge contents sharing.

Although there has been widespread use of web based e-learning applications for distance and classroom learning [3], yet little has been done to critically examine their usability [4]. Usability must be thoroughly examined and used in making sure that students do their studies with lots of ease of learning. Therefore in suggesting the e-learning, it will also help to supplement it with face-to-face teaching and active learning as a tool for instruction [5]. Active learning which requires selection of appropriate teaching approaches such as peer teaching, will definitely help students to develop in-depth understanding and expertise on the subjects taught. [6].

## 2. E-Learning And E-Teaching

According to [7], e-learning, is defined as learning and teaching online through network technologies. For e-learning initiatives to be successful, it has to either be based upon asynchronous or synchronous communication.

Conventionally, Web-based communication has relied more on asynchronous, time-delayed systems as opposed to synchronous conferencing systems which have often played a supplementary role of socializing or virtual office hours in online courses [8].

By taking advantages of the technology available, one can now access e-services through various means of technology, such as desktop computer or laptop, PDAs or cellular phones. For e-learning students to be involved in the e-learning courses they must be connected via technologies like PSTN or ISDN lines, or DSL or cable modems or from vicinity that enjoys broadband connectivity. With such varied process capabilities supported by different devices, it will create heterogeneity of client capabilities and increase the number of accessing methods [9].

The increase of bandwidth capabilities have led to the growing popularity of synchronous e-learning [10]. Synchronous e-learning or also referred in this article as e-Teaching, commonly supported by media such as videoconferencing and chat, has the potential to support e-learners in the development of learning communities. Learners and teachers experience real time environment in synchronous e-learning has more social support such as has express companionship, emotional support and can eliminate emotional distraction.

In offering e-learning facilities to prospective students, there are major factors that need to be considered so that e-learning that is been offered will become a success. As [11] has stated that some systems offer e-learning service has been successful in adapting accordingly to identify abilities, learning attitudes and preferred ways of study. [12] has also highlighted that most employed content and e-learning provision platforms are not flexible enough to dynamically meet varying user needs and connection characteristics. According to [12], while even the more sophisticated ones which usually support either personalization of content according to user profile or emphasize user-service vicinity factor, but are still lacking in taking into account all the parameters and factors that might differentiate a session from another session.

There are several problems pertaining to implementing e-learning but two issues stand out from the rest. The first is that the course material used in e-learning sometimes is unattractive and uncompelling. Secondly, is the lack of availability for pedagogies in e-learning. The pedagogies refers to the 'teachers' teaching the various chapters of a course via the e-learning process. These two issues are related to each other as well. What is perhaps surprising is that a very limited attention is being paid to the issue of pedagogies required for e-learning.

There are also many criteria that need to be considered while evaluating the e-learning course. The two most important criteria for evaluating application and quality in eLearning are that it should 'Function technically without problems across all users' and have 'clearly explicit pedagogical design principles appropriate to learner type, needs and context'. A lot of researches that has been conducted in indentifying the efficacy of the process and effective e-Learning, has indicated that online communities must be build to collect knowledge together as they work on interesting and realistic projects and problems that we come across in the life. Even to learn the robust "soft skills" (such as negotiation or sales techniques, leadership) people must compile and build upon their existing knowledge by using the new knowledge in various ways.

Teaching approaches also need to be constantly refreshed and paid attention to. If less focus is paid on teaching approaches or pedagogies in the online learning, it may lead to inactive learning styles. According to [13], main characteristics for computer technology to reproduce the teaching environment, it must possess an environment that will enable learner to be active, rich with the essential properties of what has to be learned, content should be structured and should not be complex, and should provide communication system for the knowledge to be presented. This essential element can be virtually setup using current technologies related to teleconferencing tools and broadband network infrastructure. Based on teaching environment proposed by Schneider, the main components that are involved in the environment can be concluded as consisting of three main components; teacher and students, knowledge/teaching materials and communication system as shown in Figure 1. In this case, e-Teaching tool requires functionalities to support all of these components to be interactive and communicable as described in the proposed framework.

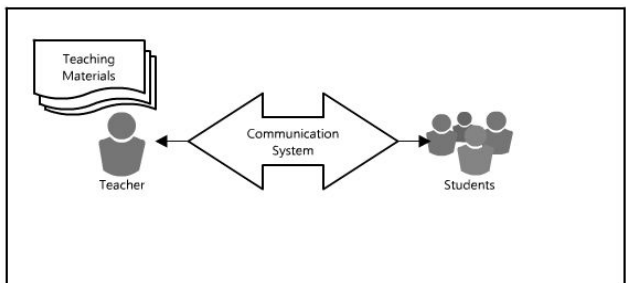

Fig. 1. Teaching environment

Additionally, [14] has also stressed that there other elements such as planning and designing of the structure, process, interaction and evaluation aspects of an online course must also be looked into. Some activities within this category might include developing curriculum materials such as creating presentations and lecture notes on the course site, and providing audio/video mini-lectures, offering a mix of individual and group activities along with a clear schedule for their completion, and providing guidelines on how to use the medium effectively.

In creating a conducive and successful e-learning mode, e-learning must be able to provide ways for teachers to engage actively with students in supporting active learning either in offline and online mode. It also should provide facilities in creating online communities. Therefore synchronous or two ways communication should be in place for online collaboration in order to really fulfill the requirement of active learning in the e-learning. Instant e-Teaching term emphasizes the implementation of e-learning tool as important medium for teachers to help student actively and effectively in their learning.

## 3. Framework of Instant E-Teaching

Instant e-Teaching framework proposes a solution for a generic e-Teaching tool that will enable live teaching session to be executed online. The framework model is designed to include some quality of software metrics as defined in ISO 9126. It consists of the internal and external attributes of e-Teaching tool such as usability (external) and efficiency (internal) of e-Teaching system.

The term "usability" refers to a set of multiple concepts, such as execution time, performance, user satisfaction and ease of learning ("learnability"), taken together [15]. Instant Messaging system such as Skype, Yahoo Messenger, Windows Live Messenger, AOL Messenger and other similar system have become very popular among all internet users. For example, registered users for Skype tool until 2010 is 560 million users [16], Windows Live Messenger until 2009 is 330 million users [17], and Yahoo Messenger until 2008 is 248 million users [18]. The figures show a high usability system indicator from the user's perspective. Thus, this framework exploits the instant messaging functionalities to be embedded into e-Teaching tool in order to gain high usability of the system.

### 3.1 Functionalities

The framework defines basic functionalities that should exist in the e-teaching environment. It mainly inherits the basic components from instant messaging such as text message, voice over IP, user list and presence status. It is also enriched with additional features in order to support teaching such as live page and pointer sharing, and navigation or command control. Table 1 summarizes the components which represents the educational setup required for an effective teaching environment.

Table 1. e-Teaching Environment

| Teaching Environment | e-Teaching Environment |
|---|---|
| Teacher/Student | User list & presence status |
| Communication System | Text message, voice over IP |
| Knowledge Material | Page / Pointer Sharing, Navigation/Command control |

a) User list & presence status

Teacher and students can be virtually presented using an iconic user list (Figure 2) to show their presence. The presence status is based on event of user login or logout activity. If a user login to the system, his icon will be displayed in full colour. Otherwise, his icon will be only displayed in grey. Presence status is also used to notify whether the student is active or inactive (idle) in the virtual classroom. If user did not make any interaction with the computer such as no mouse movement or keystroke within 5 minutes, system will notify the status of the user as idle in the user list.

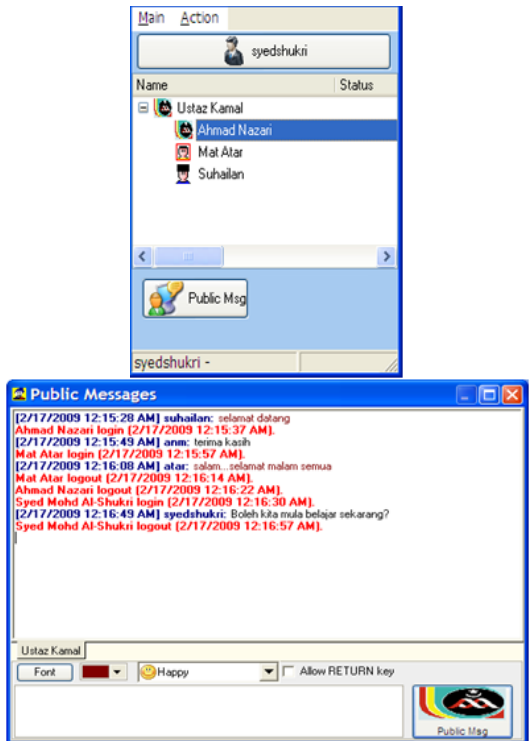

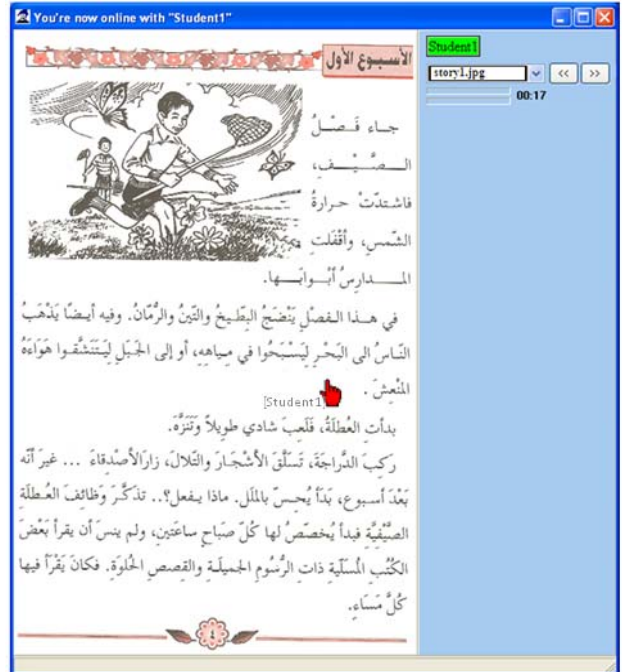

Fig. 3. Voice virtual room

Fig. 2. User list and public text message

b) Text Message and Voice Over IP

In order to communicate with each other, public text message board is provided. This instant text message can be used during the online session by all students and teacher. Voice over IP communication is accessible on one to one session when a teacher invites a student to join his private virtual classroom.

c) Page/pointer Sharing and Navigation Control

In private virtual classroom, both teacher and student will be accessing similar knowledge material page display with a live updated cursor pointer from both parties as shown in Figure 3. Teacher can alter the content page to be shown on both displays.

In term of "efficiency" of the framework model, it depends on the sub metrics of time and resource behavior. However, these sub metrics are debatable because it varies with the influences of software and hardware environment. Timing behavior is also varies with the speed of underlying system layers and also can be considered as resource behavior. Thus, instead of discussing on the time and resource behavior, [19] suggested to define "task" in the context of the application and should somehow correspond to typical tasks of the systems. Our framework model defines the arrangement of server and client tasks in Instant e-Teaching to ensure efficiency of the system.

3.2 Task Architecture

The framework defines several tasks in e-Teaching. Tasks are categorized as server's and client's tasks. Example of tasks are command handler, live e-Teaching session handler, file data manager and GUI handler.

a) Server Architecture

This system is running under client-server mode architecture. Figure 4 shows the three main components used in server application.

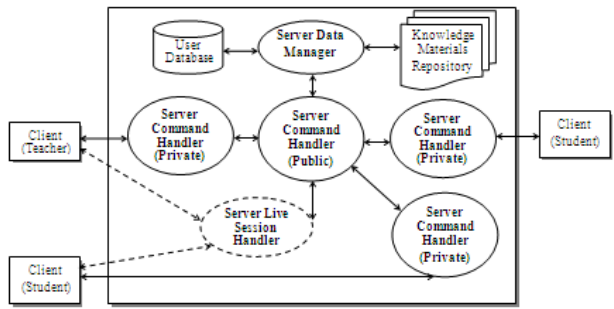

Fig. 4. Server components

i) Server Command Handler. This handler is the main thread component that provides actions and communication to the connected clients. Each client will be assigned with a dedicated private command handler and share a main public command handler. Client can request actions such as authorizing user authentication, posting public/private text message, reading and listing knowledge materials using the private command handler. While the shared public command handler enables server to communicate with the connected clients such as updating other client's status, sending alert to participate in e-Teaching session and broadcasting text messages received from other client. Server Command Handler will communicate with Server Data Manager component in order to response to the client request such as uploading knowledge material to the client. If client requests for live e-Teaching session, this module will: -

• Create a dedicated thread called as Server Live Session Handler to provide a one-to-one voice over IP and live information sharing between two respective clients. This handler will be automatically deleted when the session ended by the clients
• Interact with Server Data Manager to get current personalize knowledge material page for the client and post the page (in image format) between the clients.

ii) Server Live Session Handler. This handler creates dedicated voice communication connection at server side that links two respective clients. Live session handler will also interact with server command handler to request posting of knowledge material page to both clients. Thus client will get same page display while communicating using voice over IP. It also enables sharing live update of current mouse pointer on the page of the two clients. The mouse pointer is continuously pointed by clients to show where the client is pointing to in the page and updated to both parties on every second.

iii) Server Data Manager. This handler manages all the requests from command handler to read, write, update or delete user information in database and files in knowledge material repository. This thread manages concurrency access requested from Server Command Handler. It also enhances the reliability of the Server Command Handler by separating the Data Manager thread in handling secondary storage in term of speed and file size transfer.

b) Client Architecture

Figure 5 shows the four main components involved in the server application; Client Command Handler, Client Live Session Handler, GUI Manager and Client Data Manager.

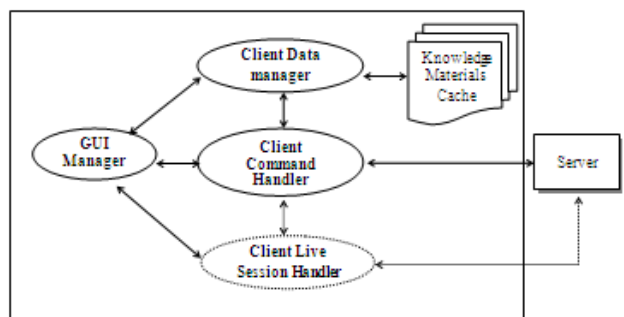

Fig. 5. Client's components

i) Client Command Handler. This command handler is main communication thread that creates synchronous link between client and server. It enables client to post actions to the public Server Command Handler and receives responses from server. It will also listen to the private Server Command Handler for any pushing content coming from server such as receiving other client public text messages. If this component receives knowledge material page (file) from server, it will communicate with Client Data Manager to check whether the file is already cached in the client local storage before proceed to receive and store the file locally. This component will interact with GUI manager to display whatever information received from server and execute specific actions required from server such as popup of voice virtual classroom window.

ii) Client Live Session Handler. This handler creates a voice communication connection at client side that links to the server. Current cursor coordinate will be also submitted to the server to be updated to the other client. It will pass information of the cursor pointer location from the other client and the current selected knowledge material page information to GUI Manager.

iii) GUI Manager. This component will display all the information received from the other three client components to its own windows as described below: -

- Client Live Session Handler - cursor coordinates in the voice virtual classroom window and selected knowledge material page.
- Client Command Handler – user list, user status, and text messages in the public text messages window
- Client Data Manager - selected knowledge material file in the voice virtual classroom window. This file is sent to GUI Manager after it has requested the page content based on the knowledge material page information given by Client Live Session Handler.

iv) Client Data Manager. This handler manages all the knowledge material files received from server when the Client Live Session Handler gets started. It will store the files into local storage at the client side and can reject from receiving file from server if the file is already up-to-date. Meanwhile, this component will pass the file contents to GUI Manager to be displayed after receiving request from GUI Manager.

## 4. Experiment

The prototype of Instant e-Teaching framework has been tested with 30 correspondents with IT background, who act as students during the session. Each student need to sit for a 10 minute teaching and learning session with a lecturer using the tool. All students are required to login to the system and get interact each other using public text messages chat room. A student will be called one by one to start a teaching and learning session with a dedicated teacher online. An image page containing the content of the current knowledge material is popup to the student screen after the teacher invited the student. Teacher starts to communicate and teach the student via voice over IP and use the pointer to show which part of the image is been referred to. Student is able to response instantly with the teacher via the same voice over IP tunnel and can also point his pointer to be shown at teacher's screen. The student will get some explanations and generate discussion with the teacher based on the image notes shown on screen. After 10 minutes experience using the tool, student will have to answer questionnaires.

### 4.1 Result

Overall, usability of the tool in e-teaching for adult is given positive comments. 87% of users have remarked that the tool has save lot of time in accessing learning material and engaging face to face lectures. They can easily learn directly from an instructor by asking questions that might need more explanation.

Emotional distraction in learning was also eliminated as student can privately attend the course lectures in virtual environment. Emotional distraction here refers to a feeling of uneasiness and ashamed of doing mistakes with peers realizing the doer making those mistakes. There were 90% of users have concurred that they are no longer distracted from emotional factors when using the prototype. Although names have appeared in the global user list of the online communities, students feel comfortable because they were sitting at their own physical room and no one is focusing on his/her action in that virtual classroom.

Meanwhile, 90% of users claimed that the use of ICT elements in the prototype is very helpful in executing the e-Teaching. The use of voice over IP really helped student to communicate with teacher to gain synchronous and instant responses. Active engagement was created among them enabling live teaching and learning process to be executed. Although some of the participants felt the both side of live cursor pointer quite confusing, but eventually they found that the pointer is really assisting them to support voice command in teaching and learning environment. A majority of users were able to complete the learning session in 10 minutes without any problems although they having never seen and used the program before. And most of them were actively communicate among themselves via instant public chatting without any prior instruction or tutorial given.

They agreed that the tool did annotate the real teaching environment and can replace the traditional classroom. This is a good indication of a highly intuitive interface and speaks highly of the software's potential as an educational tool. It is also encouraging that almost all participants reported that the tool is suitable to be used in teaching other common lecture. Other examples of common lectures are languages practices, religious sermon, theoretical and terms discussion, and laws subject.

### 4.2 Discussion

This framework focuses in creating a working model of e-Teaching tool by considering the usability and efficiency of the systems. In order to satisfy the usability and friendliness of the e-Teaching tool, all the actions must be designed to be simple and fast access. Simplicity can be achieved by minimizing procedural steps and fast access by including automated procedures. One of the examples that realize the principles is the knowledge material repository which is used to enable teacher to upload notes without mush hassle. While in a live session of e-Teaching, teacher can just upload an image file and it will automatically appear as teaching page with the student. At the same time the file will be stored in the knowledge material repository as students' references. Using voice communication, it will become the most immediate, effective and easy tool to execute teaching and learning supporting the page notes. Meanwhile, instant text message feature enables students to get instant responses among themselves and also teacher.

Most of teleconferences or instant messaging systems that apply live and real time communication elements involving text, image, voice, video or desktop sharing will experience some performance issues. Such common issues are reliability and availability issue. These include system responses delay, crash, hang and unresponsive system. Thus designing such system that involves components used in e-Teaching tool requires a very experience programmer. Through the framework, dedicated components are used to solve system reliability. Threads are used to handle concurrent actions and also to avoid system to be totally crashed. For example, if the system crashed at the client data manager component due to network data loss, user can still communicate each other via voice or text messages in the client live session handler or client command handler components. At the same time, it also increases the availability of the system.

## 5. Conclusions

E-learning tools have been widely implemented in academic institution in providing easy long life access and self-learning system for students. However, instead of focusing on the online learning (e-Learning) which does not easily promote an active learning, an instant e-Teaching model was highlighted here as usable and efficient in supporting distance learning course. Functionalities that enable such peer teaching to provoke active learning are recommended in the framework. Meanwhile, the efficiency of the framework was elaborated by assigning tasks as generic and specific custodian in executing all the functions needed in online teaching.

## Acknowledgments

This research project was funded by Universiti Sultan Zainal Abidin, Malaysia since 2008 has led to the realization of modeling generic tool for usable and efficient online teaching environment.

**Suhailan Safei** is currently working as Deputy Dean, Faculty of Informatics, Universiti Sultan Zainal Abidin, Malaysia. He has 3 years of teaching experience and 7 years of working in real time software development. He has obtained his B.Sc. in Computer Science and M.Sc. Computer Science (Real Time Software Engineering) from Universiti Teknologi Malaysia.

**Mat Atar Mat Amin** is currently working as Head of Department, Faculty of Informatics, Universiti Sultan Zainal Abidin, Malaysia. He has 14 years of teaching experience and 3 years working experience in University's Research Management Center.

**Ahmad Nazari Mohd Rose** is currently working as a senior lecturer, Faculty of Informatics, Universiti Sultan Zainal Abidin, Malaysia. He has 20 years of teaching experience and working experience in Shell oil company.

**Mohd Nordin Abdul Rahman** is currently working as Associate Prof. and Dean of Faculty of Informatics, Universiti Sultan Zainal Abidin, Malaysia. He has more than 15 years of teaching and has published numbers of papers in referred international journals.